\documentclass[preprint,preprintnumbers,amsmath,amssymb]{revtex4-1}

\usepackage{caption}
\usepackage{graphicx}
\usepackage{subfigure}

\begin{document}

\title{Accurate Force Field for Molybdenum by Machine Learning Large Materials Data}
\author{Chi Chen}
\affiliation{Department of NanoEngineering, University of California San Diego, 9500 Gilman Dr, Mail Code 0448, La Jolla, CA 92093-0448, United States}
\author{Zhi Deng}
\affiliation{Department of NanoEngineering, University of California San Diego, 9500 Gilman Dr, Mail Code 0448, La Jolla, CA 92093-0448, United States}
\author{Richard Tran}
\affiliation{Department of NanoEngineering, University of California San Diego, 9500 Gilman Dr, Mail Code 0448, La Jolla, CA 92093-0448, United States}
\author{Hanmei Tang}
\affiliation{Department of NanoEngineering, University of California San Diego, 9500 Gilman Dr, Mail Code 0448, La Jolla, CA 92093-0448, United States}
\author{Iek-Heng Chu}
\affiliation{Department of NanoEngineering, University of California San Diego, 9500 Gilman Dr, Mail Code 0448, La Jolla, CA 92093-0448, United States}
\author{Shyue Ping Ong}
\email{ongsp@eng.ucsd.edu}
\affiliation{Department of NanoEngineering, University of California San Diego, 9500 Gilman Dr, Mail Code 0448, La Jolla, CA 92093-0448, United States}

\begin{abstract}
In this work, we present a highly accurate spectral neighbor analysis potential (SNAP) model for molybdenum (Mo) developed through the rigorous application of machine learning techniques on large materials data sets. Despite Mo's importance as a structural metal, existing force fields for Mo based on the embedded atom and modified embedded atom methods do not provide satisfactory accuracy on many properties. We will show that by fitting to the energies, forces and stress tensors of a large density functional theory (DFT)-computed dataset on a diverse set of Mo structures, a Mo SNAP model can be developed that achieves close to DFT accuracy in the prediction of a broad range of properties, including elastic constants, melting point, phonon spectra, surface energies, grain boundary energies, etc. We will outline a systematic model development process, which includes a rigorous approach to structural selection based on principal component analysis, as well as a differential evolution algorithm for optimizing the hyperparameters in the model fitting so that both the model error and the property prediction error can be simultaneously lowered. We expect that this newly developed Mo SNAP model will find broad applications in large-scale, long-time scale simulations.
\end{abstract}

\maketitle
\section{Introduction}

One of the fundamental challenges in computational materials science is the trade-off between accuracy and scale (length/time). \textit{Ab initio} or first principles methods, such as those based on the density functional theory (DFT) formalism\cite{Kohn1965,Sham1983}, have become the method of choice in problems where good chemical accuracy is required. However, due to the high computational cost of solving the Schr\"{o}dinger equation, most \textit{ab initio} calculations are limited to $< 1000$ atoms, and \textit{ab initio} molecular dynamics (AIMD) simulations are limited to a time scale of hundreds of picoseconds. Alternatively, simulations of materials can be carried out using empirical force fields\cite{Senftle2016,Rappe1992}, which assume an explicit functional form for the relationship between atomic configurations and local energies. Force field calculations are usually orders of magnitude faster than \textit{ab initio} calculations, allowing for simulations of systems that contain thousands or even millions of atoms, and over time scales of nanoseconds to microseconds. However, most empirical force fields lack transferability and fail in simulations of complex chemistry with significant variations in chemical bonding and local environments.

In recent years, there have been an upsurge in the application of machine learning (ML) methods to physics problems and material simulations\cite{Zdeborova2017,Carleo2017,Carrasquilla2017,Ulissi2017,Jain2016a,Chen2015,Chen2017}. ML models are often trained on DFT calculations, aiming at reproducing material properties with DFT-level accuracy. Notable successes have been demonstrated in predicting atomization energies\cite{Hansen2015}, highest occupied molecular orbital and lowest unoccupied molecular orbital eigenvalues,\cite{Montavon2013} dielectric constants,\cite{Pilania2013} energies\cite{Meredig2014}, etc.\cite{Olivares-Amaya2011} Similarly, there have been attempts to construct force field models using ML. The models include high-dimensional neural networks\cite{Behler2007}, gaussian approximation potentials\cite{Bartok2010}, kernel ridge regression\cite{Botu2015}, and momentum tensor potentials\cite{Shapeev2016}. A typical ML model works by first converting structures into numerical values, called features or descriptors, and then the model directs the features into a training procedure along with DFT-calculated quantities, such as energy, force and stress, as the model targets.

A critical component in the ML model development is the choice of feature functions. Ideally, the feature conversion should be invariant to transformations that preserve material properties, for example, permutation of equivalent atoms, rotations and translations. Some examples of proposed feature functions that satisfy or partially satisfy these invariant properties include the Coulomb matrix and its derivatives\cite{Rupp2012b,Moussa2012}, bag-of-bonds\cite{Hansen2015}, symmetry functions\cite{Behler2011}, bispectrum coefficients\cite{Bartok2010}, among others.\cite{Bartok2013a, Pham2017} In principle, ML models can be systematically improved with more training data, if the feature functions are complete and can distinguish unique local environments. Atomic distance-based features such as Weyl matrices\cite{Weyl2016} and the histogram of pair distance distributions\cite{Swamidass2005} cannot fulfill these requirements because they fail to provide a unique representation of the local environment\cite{Bartok2013a}.

Recently, \citet{Bartok2010} introduced the bispectrum coefficients as a means of mapping the local atomic density function into invariant representations. The bispectrum coefficients have the advantage of providing an almost one-to-one representation of the atomic neighborhood. Subsequently, Thompson et al.\cite{Thompson2015,Wood2017} demonstrated that a spectral neighbor analysis potential, or SNAP, that expresses energies, forces and stress tensors as a linear model of the bispectrum coefficients and their first derivatives can produce quantum-accurate property predictions for Ta and W. One of the key advantages of the SNAP formalism is that the DFT energies, forces and stress tensors can be trained in the same framework. Furthermore, due to its simple formalism, the model is less likely to experience overfitting compared to conventional force fields such as the embedded atom model (EAM) or modified embedded atom model (MEAM), which usually require the optimization of nested nonlinear functions.

In this work, we will present a systematic ML approach to build a SNAP model for Mo. Mo is one of the most important structural metals, valued for its ability to withstand high temperatures, high corrosion resistance, and excellent strength-to-weight ratio. Despite its importance, currently available force fields for Mo based on the EAM\cite{Zhou2004} and MEAM\cite{Park2012} still do not provide satisfactory accuracy on many properties. This work builds on the excellent work of \citet{Bartok2010} and Thomson et al.\cite{Thompson2015,Wood2017}, but improves on the training procedure in two ways. First, we outline a  principal component analysis approach to the selection of training structures, which are obtained from large diverse DFT datasets that we have accumulated via our previous work on Mo grain boundaries\cite{Tran2016a} and surfaces\cite{Tran2016}, supplemented with additional data obtained via high-throughput DFT calculations of liquid and solid structures. Second, we propose the use of a differential evolution algorithm to simultaneously optimize the hyperparameters, such as the cutoff radius and weights of the training structures, and the model parameters. We demonstrate that this machine-learned Mo SNAP model can achieve near-DFT accuracy across a wide range of properties, including energies, forces and stress tensors, elastic properties, melting point, surface and grain boundary (GB) energies, outperforming currently available potentials for Mo.

\section{Bispectrum and SNAP formalism}

The bispectrum and SNAP formalism has been covered extensively in previous works.\cite{Bartok2010,Thompson2015} We will only provide a brief summary of the key concepts here, and refer interested readers to those excellent works.

The basic idea of the bispectrum formalism is to map a 3D local atomic neighbor density into a set of coefficients that satisfy the invariant properties. The atomic neighbor  density around atom $i$ at location $\mathbf{r}$ is expressed as:
\begin{equation}
\rho_i(\mathbf{r}) = \delta(\mathbf{r}) +\underset{r_{ii^\prime}<R_c} \sum f_c(r_{ii^{\prime}}) w_{i^\prime} \delta(\mathbf{r- r}_{ii^\prime})
\end{equation}
where $i^\prime$ denotes a neighbor atom, and $w_{i^\prime}$ is the dimensionless weight to distinguish atom types. The weight is set as 1 in this work as only one element is present. The cutoff function $f_c(r)$ ensures that the neighbor atomic density goes smoothly to zero when the distance $r_{ii^\prime}$ is greater than the cutoff radius $R_c$.

The angular information in the 3D local density function can be projected onto spherical harmonic functions $Y_m^l(\theta, \phi)$. In the bispectrum approach, the radial component is converted into a third polar angle defined by $\theta_0 = \theta_0^{max}\frac{r}{R_c}$. Thus the density function can be represented in the 3-sphere $(\theta, \phi, \theta_0)$ coordinates instead of $(\theta, \phi, r)$. The density function defined on the 3-sphere can then be expanded using 4D hyperspherical harmonics as follows:

\begin{equation}
\rho(\mathbf{r}) = \sum\limits_{{j=0,\frac{1}{2},...}}^{\infty} \sum\limits_{{m=-j}}^{j} \sum\limits_{{m^\prime=-j}}^{j} u_{m, m^\prime}^jU_{m, m^\prime}^j(\theta, \phi, \theta_0 )
\end{equation}
where the coefficients $u_{m, m^\prime}^j$ are obtained as the inner products between the density function and the basis, given by the following:

\begin{equation}
u_{m, m^\prime}^j = U_{m, m^\prime}^j(0, 0, 0) + \underset{r_{ii^\prime}<R_c} \sum f_c(r_{ii^{\prime}}) w_{i^\prime} U_{m, m^\prime}^j(\theta, \phi, \theta_0 )
\end{equation}

The bispectrum coefficients $B_{j_1, j_2, j}$ can then be obtained via the following equation:
\begin{equation}
B_{j_1, j_2, j}= \sum\limits_{m_1, m_1^\prime = -j_1}^{j_1} \sum\limits_{m_2, m_2^\prime = -j_2}^{j_2}  \sum\limits_{m, m^\prime = -j}^{j} \left( u_{m, m^\prime}^j \right )^\text{*} H^{\substack{jmm^\prime\\j_1m_1m_1^\prime}}_{j_2m_2m_2^\prime}  u_{m_1, m_1^\prime}^{j_1}
u_{m_2, m_2^\prime}^{j_2}
\end{equation}

where the constants $H^{\substack{jmm^\prime\\j_1m_1m_1^\prime}}_{j_2m_2m_2^\prime}$ are coupling coefficients and $||j_1-j_2|| \leq j \leq ||j_1 + j_2||$.

In the SNAP formalism, the energy $E_{SNAP}$, force $\boldsymbol{F}_{SNAP}^j$ and stress $\boldsymbol{\sigma}_{SNAP}^j$ are related to the bispectrum coefficients $\boldsymbol{B}$ by the following
\begin{subequations}\label{fitting_equation}
\begin{align}
E_{SNAP} & = \beta_0 N  + \boldsymbol{\beta}\cdot \sum\limits_{i=1}^N \boldsymbol{B}^i  \label{energy_equation}\\
\boldsymbol{F}_{SNAP}^j & = - \boldsymbol{\beta}\cdot \sum\limits_{i=1}^N \frac{\partial \boldsymbol{B}^i}{\partial \boldsymbol{r}_j}  \label{force_equation}\\
\boldsymbol{\sigma}_{SNAP}^j & = - \boldsymbol{\beta}\cdot \sum\limits_{j=1}^N \boldsymbol{r}_j \otimes \sum\limits_{i=1}^N \frac{\partial \boldsymbol{B}^i}{\partial \boldsymbol{r}_j}   \label{stress_equation}
\end{align}
\end{subequations}

where $\beta_0$ and the vector $\boldsymbol{\beta}$ are the coefficients in the linear models and are fitted from the DFT data to relate $E_{SNAP}$, $\boldsymbol{F}_{SNAP}$ and $\boldsymbol{\sigma}_{SNAP}$, to the structural bispectrum coefficients $\boldsymbol{B}$ and their derivatives $\frac{\partial \boldsymbol{B}}{\partial \boldsymbol{r}}$.

\section{Potential development}

\begin{figure}[!htb]
\includegraphics[width=0.8 \textwidth]{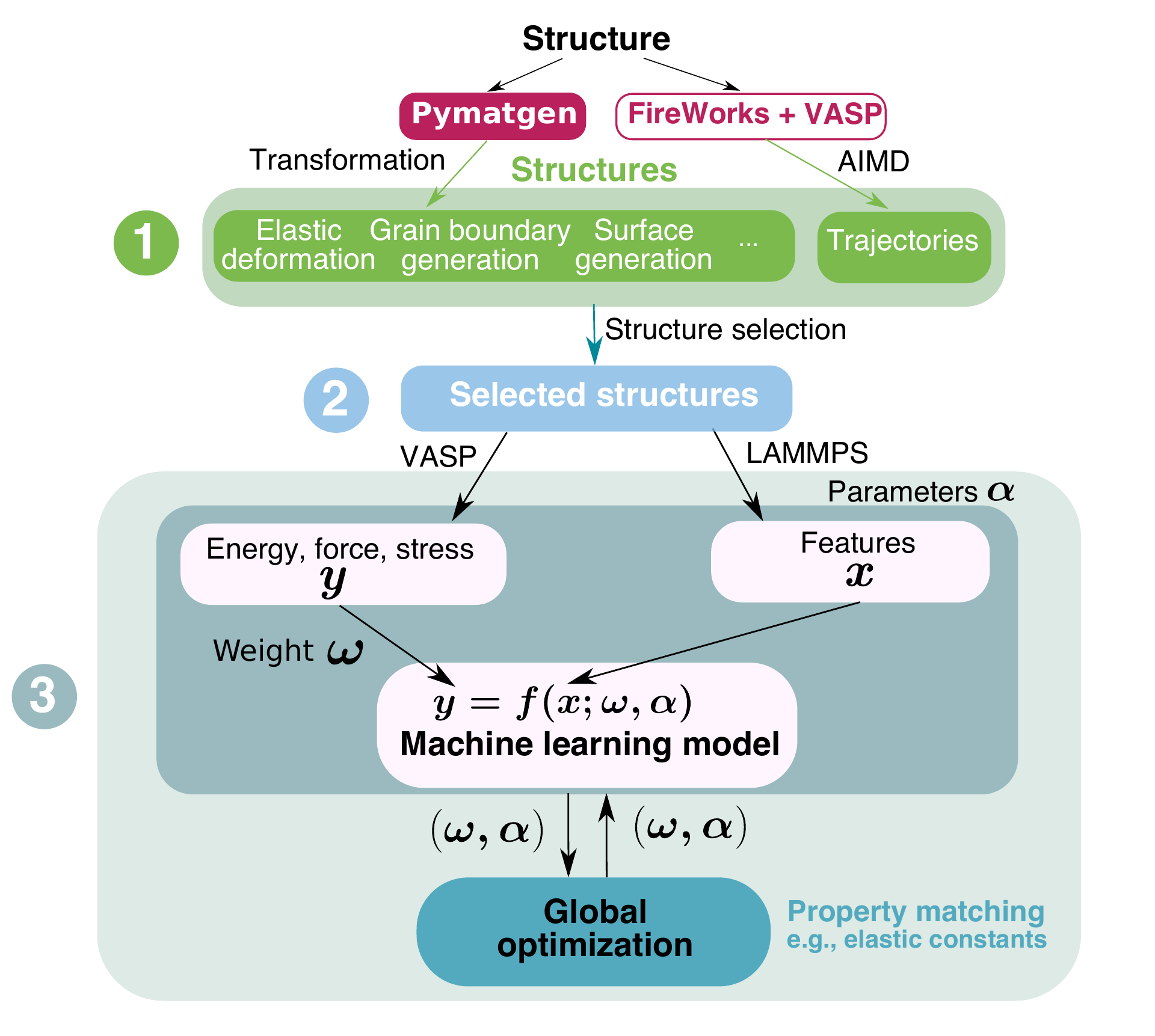}
\caption{\label{fig:illustration}  Model fitting workflow. }
\end{figure}

Figure \ref{fig:illustration} provides an overview of the potential fitting workflow, which comprises three key steps. First, a set of training structures were generated using structural transformation functions in the Python Materials Genomics (pymatgen) library,\cite{Ong2013a} as well as AIMD simulations. Second, we propose a structure selection process using principal component analysis (PCA) to identify a reasonable set of structures that provide a good coverage of the feature space for training. Third, the selected training structures were converted into bispectrum coefficients (the feature set), and DFT calculations were performed using these structures. The features and DFT results were fed into the inner loop ML model. On top of the ML model, a differential evolution global optimization algorithm\cite{Neri2010} was used to tune the weights from different data groups and the parameters used in feature calculations so that the final model can provide good predictions on material properties as well as basic quantities such as energies, forces, and stress tensors. The overall fitting can be seen as an alternating two-step process. In the inner loop, fitting of the ML model was performed. In the outer loop, the ML model generated in each iteration was then used to compute material properties such as the elastic tensors, and the differences between the predicted and reference values were then used to optimize the hyperparameters. This iterative process was continued until satisfactory accuracy was achieved for both material properties and basic quantities.

\subsection{Training data generation}

To develop an effective and robust potential, it is critical that the training data is diverse. Here, we exploit three sets of pre-computed data from our previous work:

\begin{enumerate}
  \item Ground state structures and energies for Mo from the Materials Project database.\cite{Jain2013}
  \item Surface slab structures from the Crystalium database,\cite{Tran2016, crystalium2016} which contains the pre-computed surface energies and Wulff shapes of most elements in the Periodic Table. For Mo, the data on all 13 distinct surfaces up to a maximum Miller index of three - (100), (110), (111), (210), (211), (221), (310), (311), (320), (321), (322), (331), and (332) - were included.
  \item GB structures from our previous study of the effect of dopants on Mo GBs\cite{Tran2016a}. Specifically, DFT data from the relaxation of the (100) $\Sigma 5$ twist and  (310) $\Sigma 5$ tilt, and static calculations of (110) $\Sigma 3$ twist, (111) $\Sigma 3$ tilt and (110) $\Sigma 11$ twist boundaries were included.
\end{enumerate}

We then further augmented this dataset with additional structures incorporating elastic, defect, dynamics and phase transformation information, as follows:

\begin{enumerate}
  \item Strains of -10\% to 10\% at 1\% intervals were applied to a $3\times3\times3$ supercell containing 54 Mo atoms in six different modes, as described in the work by De Jong et al.\cite{DeJong2015}
  \item $NVT$ AIMD simulations of a 54-atom supercell were performed at 300 K, 3000 K and 6000 K. 40 snapshots were obtained from each AIMD simulation at intervals of 0.1 ps.
  \item An $NpT$ simulation at 6000 K was performed using a 54-atom supercell to obtain the liquid phase of Mo. Nine snapshots were extracted for DFT calculations and further deformations were also carried out on one liquid supercell structure to obtain an additional 40 training structures.
  \item AIMD simulations were performed for vacancy-containing structures at 300 K, 3000 K and 6000 K, and 40 snapshots were extracted from each simulation.
\end{enumerate}

DFT calculations on all structures were carried out using the Vienna \textit{Ab initio} Simulation Package (VASP)\cite{Kresse1996} within the projector augmented wave approach.\cite{Blochl1994} The Perdew-Burke-Ernzerhof (PBE)\cite{Perdew1996a} generalized gradient approximation (GGA) was adopted for the exchange-correlation functional, and the pseudopotential used was Mo\_{pv} 04Feb2005 with 4\textit{p}, 5\textit{s} and 4\textit{d} electrons. The kinetic energy cutoff was set to 520 eV and $k$-point density was at least 3000 per reciprocal atom. The electronic energy and atomic force components were converged to within $10^{-5}$ eV and 0.02 eV/\AA, respectively. We found that the energy error converged to less than 1 meV/atom using this scheme. For previously relaxed structures, static calculations with the same settings as current work were performed to ensure consistency. The AIMD simulations were performed with a single $\Gamma$ $k$-point and were non-spin-polarized. However, the energy, force and stress computations carried out on the snapshots were performed using the same parameters are the rest of the data. All structure manipulations and analysis of DFT computations were performed using pymatgen and automation of calculations were carried out using the FireWorks software.\cite{Jain2015a} The structures and the corresponding DFT computed data are provided at https://github.com/materialsvirtuallab/snap.

The training structures were converted into bispectrum coefficients (the features) using the implementation in the LAMMPS software \cite{Plimpton1995} by \citet{Thompson2015} Based on extensive benchmarks carried out over our dataset, we found that an order of three for the bispectrum ($j_{max} = 3$) is sufficient, in line with previous works.\cite{Thompson2015, Bartok2010} We have also kept the angle conversion factor $\theta_0^{max}$ at the default value of $0.99363 \pi$ because we do not expect it to have a significant impact on the model performance. The cutoff radius $R_c$ was further optimized during training (see later section).

\subsection{Data selection}

\begin{figure}[!htb]
\begin{center}
\subfigure[Bispectrum coefficients]{\label{fig:snapca}
\includegraphics[width=.45\textwidth]{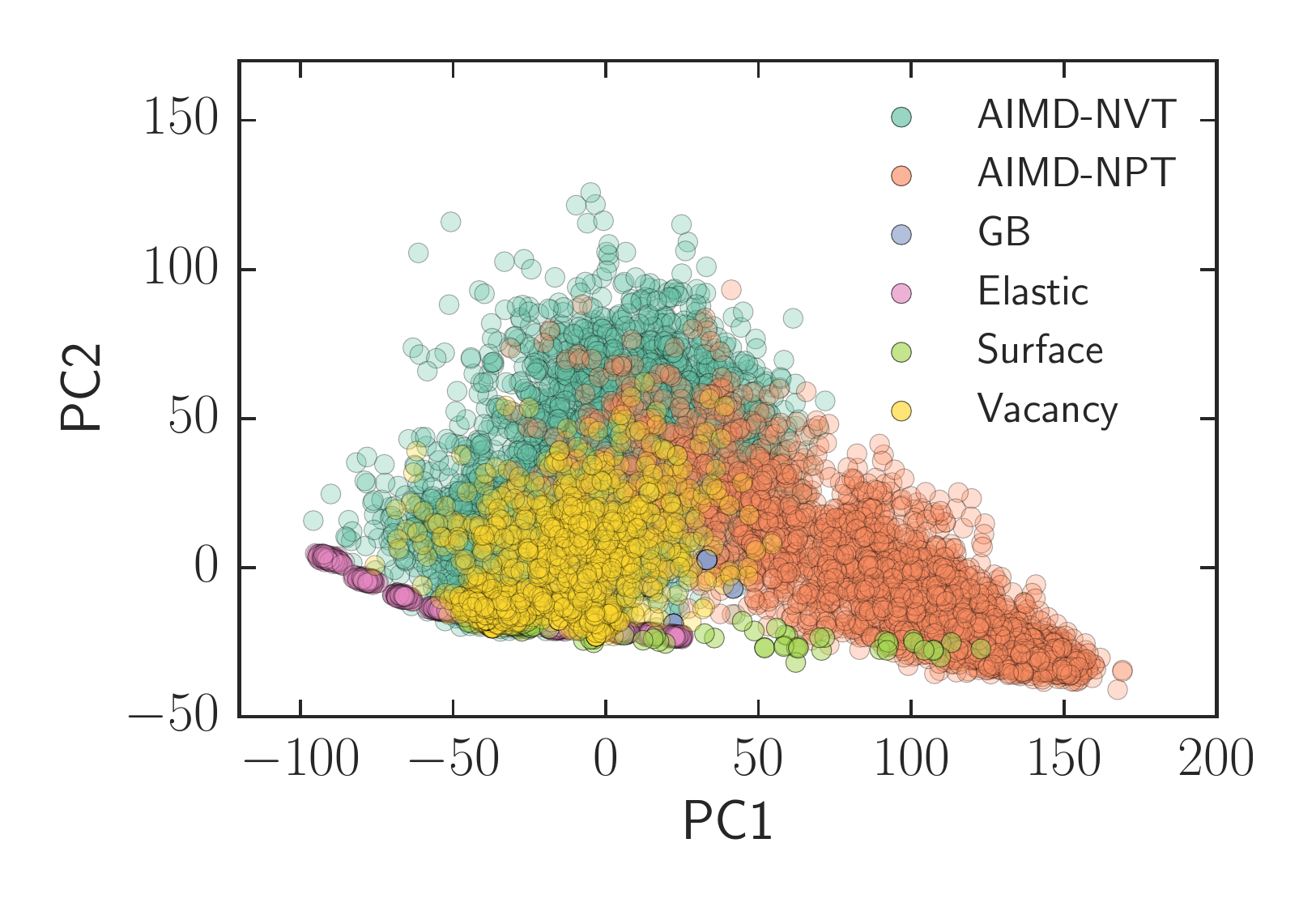}
}
\quad
\subfigure[First derivative of bispectrum coefficients]{\label{fig:snadpca}
\includegraphics[width=.45\textwidth]{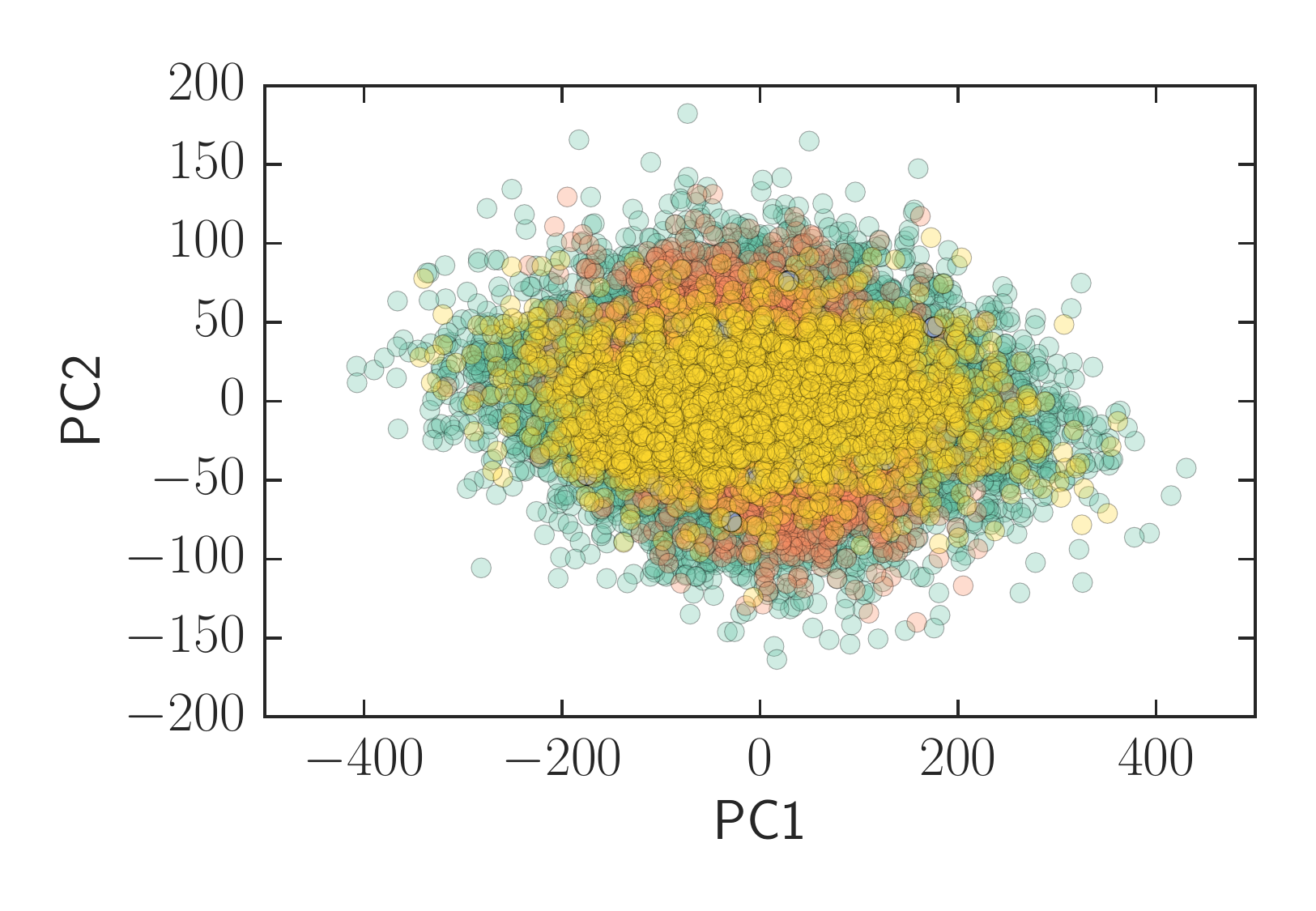}
}
\caption{\label{fig:pca} Two-dimensional projection of the principal components of (a) the atomic bispectrum coefficients and (b) their first derivatives.}
\end{center}
\end{figure}

Prior to model training, we performed an exploratory data analysis to examine the distribution of features in the feature space. The aim of this step, which was absent in prior works, is (a) to ensure that we have a good coverage of the feature space of interest, and (b) to minimize the number of structures in (relatively) expensive DFT computations. Data reduction is also particularly important for non-parametric models, e.g. kernel ridge regression, which tends to scale poorly (usually $O(n^3)$ or more) with training data size. Figure \ref{fig:pca} shows the results of PCA carried out to project the bispectrum features of the entire dataset and their first derivatives onto a two-dimensional plane formed by the first two dominating principal components (PCs). We may observe that the the bispectrum coefficients of atoms in the AIMD structures cover a wide swath of feature space. While the elastic, surface and GB data contribute additional features at the edges of the space, the features from the vacancy data sets lie within the AIMD data group. We surmise that the reason is because the AIMD structures already include a rich variety of local environments from both liquid and solid structures, some of which resemble the vacancy structures. Based on this analysis, we have excluded the vacancy dataset from the model training. We will discuss the effect of this exclusion in the Discussion section.

\subsection{Model training and optimization}

The Mo SNAP model was trained using the bispectrum coefficients as the features, and the DFT energies, forces and stress components as the outputs. Having excluded the vacancy data based on the PCA analysis, the features and DFT data were obtained from the AIMD, elastic, surface and GB structures. In line with the previous work by \citet{Thompson2015}, the DFT energies were normalized by the number of atoms in the corresponding structures, while the model-predicted virial stresses according to Equation \ref{stress_equation}, having a unit of energy, were normalized using the structural volume. The inner loop fitting of the model coefficients was done with the ordinary least squares algorithm implemented in the scikit-learn package\cite{Pedregosa2011}.

The data weights, along with the cutoff radius for $R_c$ for the calculation of features, were treated as hyperparameters for optimization. Whereas \citet{Thompson2015} relied on the DAKOTA toolkit\cite{Adams2006}, this work utilizes the differential evolution algorithm\cite{Neri2010} implemented in the widely available SciPy\cite{Jones2014} package to optimize the hyperparameters, with the target being to minimize the error between the DFT and SNAP predicted elastic constants. The lower and upper bounds for the energy weights and force weights were set as (0.5, 3000) and (0.001, 100), respectively for the AIMD, surface and GB datasets. Virial stresses were not used for these three training data groups. For the elastic data group, the bounds for energy weights and stress weights were (0.05, 10000) and (0.001, 10), respectively, while the forces were not used. The higher upper bounds for the energy and force weights for the elastic data group were chosen to ensure that the predicted energies and forces are more accurate for these groups relative to other groups. Smaller weights were chosen for the stress components, which have much large absolute magnitudes than the forces. The bounds for $R_c$ was set to (4~\AA, 5~\AA), which is up to the third nearest neighbor distance in pristine bcc Mo.

The final SNAP model coefficients ($\beta_0$ and $\boldsymbol{\beta}$ in equation \ref{fitting_equation}) are provided in Table \ref{snapcoef} below, while the optimized weights of the training data are provided in Table S1. The optimized cutoff radius $R_c$ is 4.615858 \AA, which is slightly larger than the third nearest neighbor distance in pristine bcc Mo.

\begin{table}[htp]
\begin{center}
\caption{SNAP coefficients for Mo.}
\footnotesize
\renewcommand{\arraystretch}{0.8}
\begin{tabular}{cccccc}
\hline
\hline
k \ \ & \ \  $2j_1$ \ \ & \ \ $2j_2$ \ \ & \ \  $2j$ \ \ & $\beta_k$ \\
\hline
0  &  &  &  & -17.26611796920\\
1  &  0 & 0 & 0 & 0.004313484626 \\
2  &  1 & 0 & 1 & 0.065673014215\\
3  &  1 & 1 & 2 & 0.477515651701\\
4  &  2 & 0 & 2 & 0.015311383459\\
5  &  2 & 1 & 3 & 0.775255292581\\
6  &  2 & 2 & 2 & 0.284635921402\\
7  &  2 & 2 & 4 & 0.148648095785\\
8  &  3 & 0 & 3 & 0.057357340924\\
9  &  3 & 1 & 4 & 0.192638821858\\
10  &  3 & 2 & 3 & 0.323134590663\\
11 &  3 & 2 & 5 & 0.101175999738\\
12  &  3 & 3 & 4 & 0.013745812607\\
13 &  3 & 3 & 6 & -0.032457334808\\
14 &  4 & 0 & 4 & 0.034965244240\\
15  &  4 & 1 & 5 & 0.061331627466\\
16  &  4 & 2 & 4 & 0.087925371053\\
17 &  4 & 2 & 6 & 0.118708938112 \\
18  &  4 & 3 & 5 & 0.006864594365\\
19  &  4 & 4 & 4 & -0.017509804053\\
20  &  4 & 4 & 6 & -0.017920766445\\
21  &  5 & 0 & 5 & 0.019589412881\\
22  &  5 & 1 & 6 & 0.072060016768\\
23  &  5 & 2 & 5 & 0.034468036451\\
24 &  5 & 3 & 6 & -0.035900678283\\
25  &  5 & 4 & 5 & -0.029228868192\\
26 &  5 & 5 & 6 & -0.033524823674\\
27 &  6 & 0 & 6 & 0.005947297087\\
28  &  6 & 2 & 6 & 0.075469314756\\
29  &  6 & 4 & 6 & -0.000970783042\\
30 &  6 & 6 & 6 & -0.010021282591\\

\hline
\hline
\end{tabular}
\label{snapcoef}
\end{center}
\end{table}

\section{Performance of Mo SNAP Model}

In this section, we will compare the performance of the optimized Mo SNAP model in predicting many properties of interest.

\subsection{Energies, forces and stresses}

\begin{figure}[!htb]
\begin{center}
\subfigure[Energy]{\label{fig:error1}
\includegraphics[width=.45\textwidth]{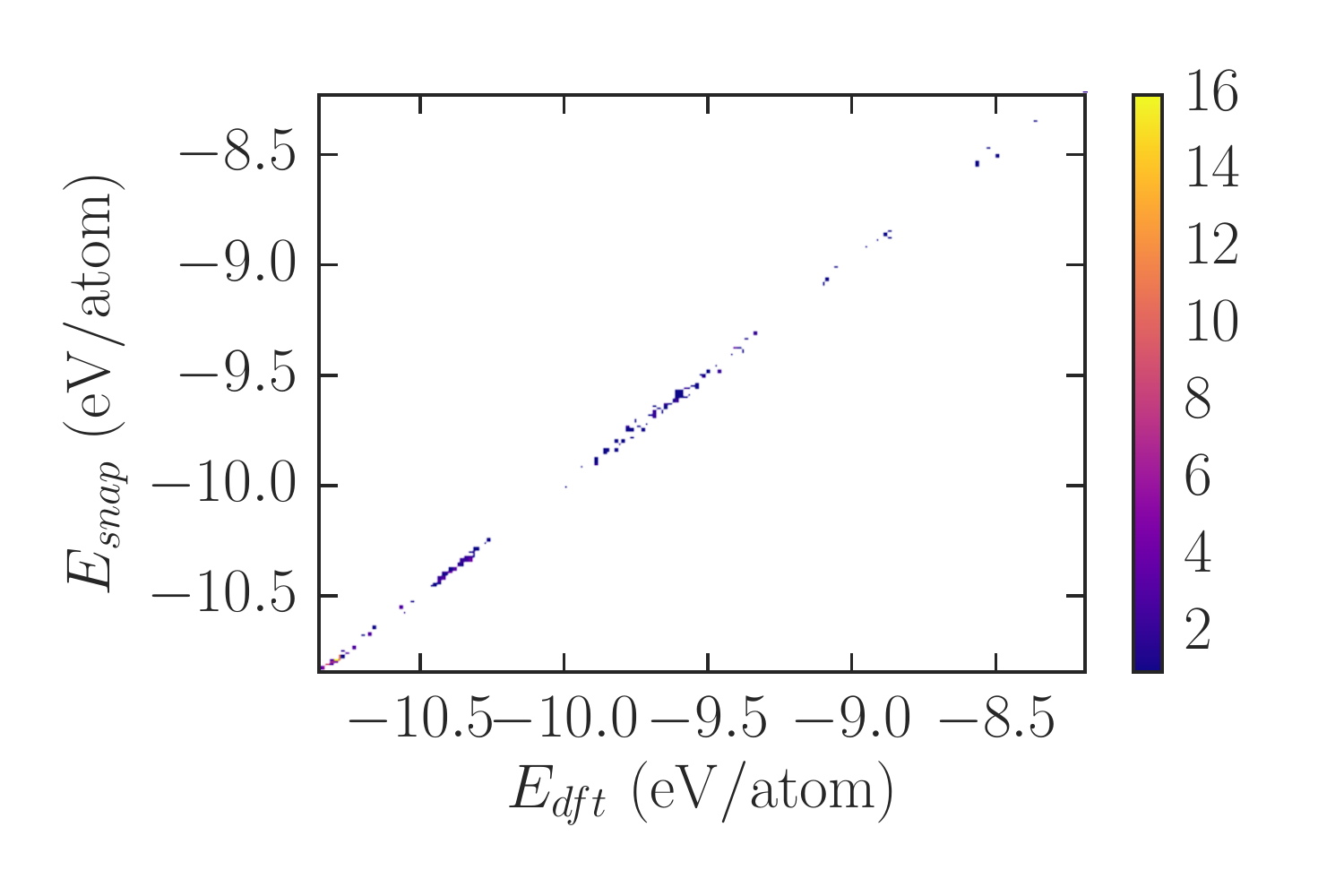}
}
\subfigure[Force]{\label{fig:error2}
\includegraphics[width=.45\textwidth]{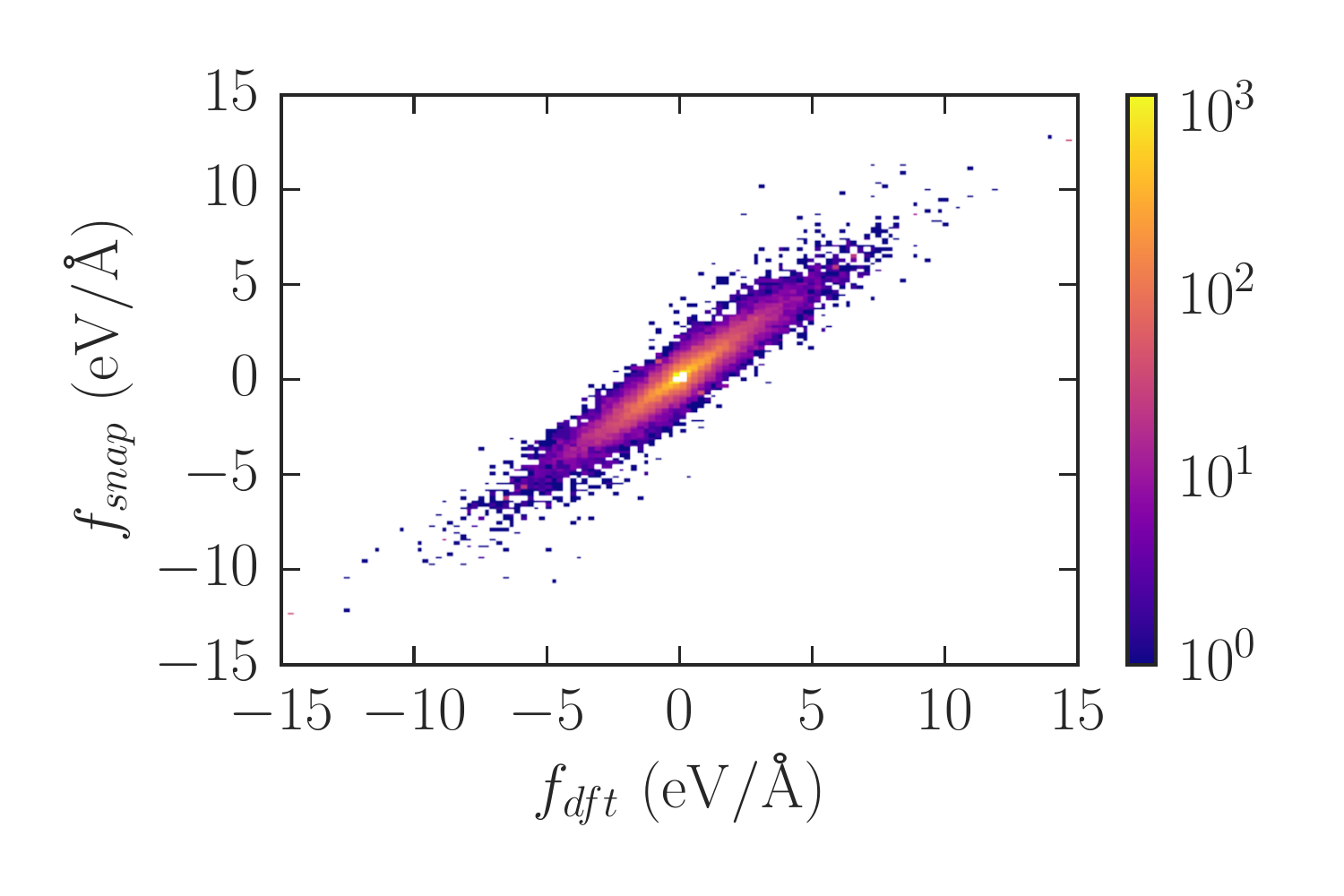}
}
\subfigure[Stress component]{\label{fig:error3}
\includegraphics[width=.45\textwidth]{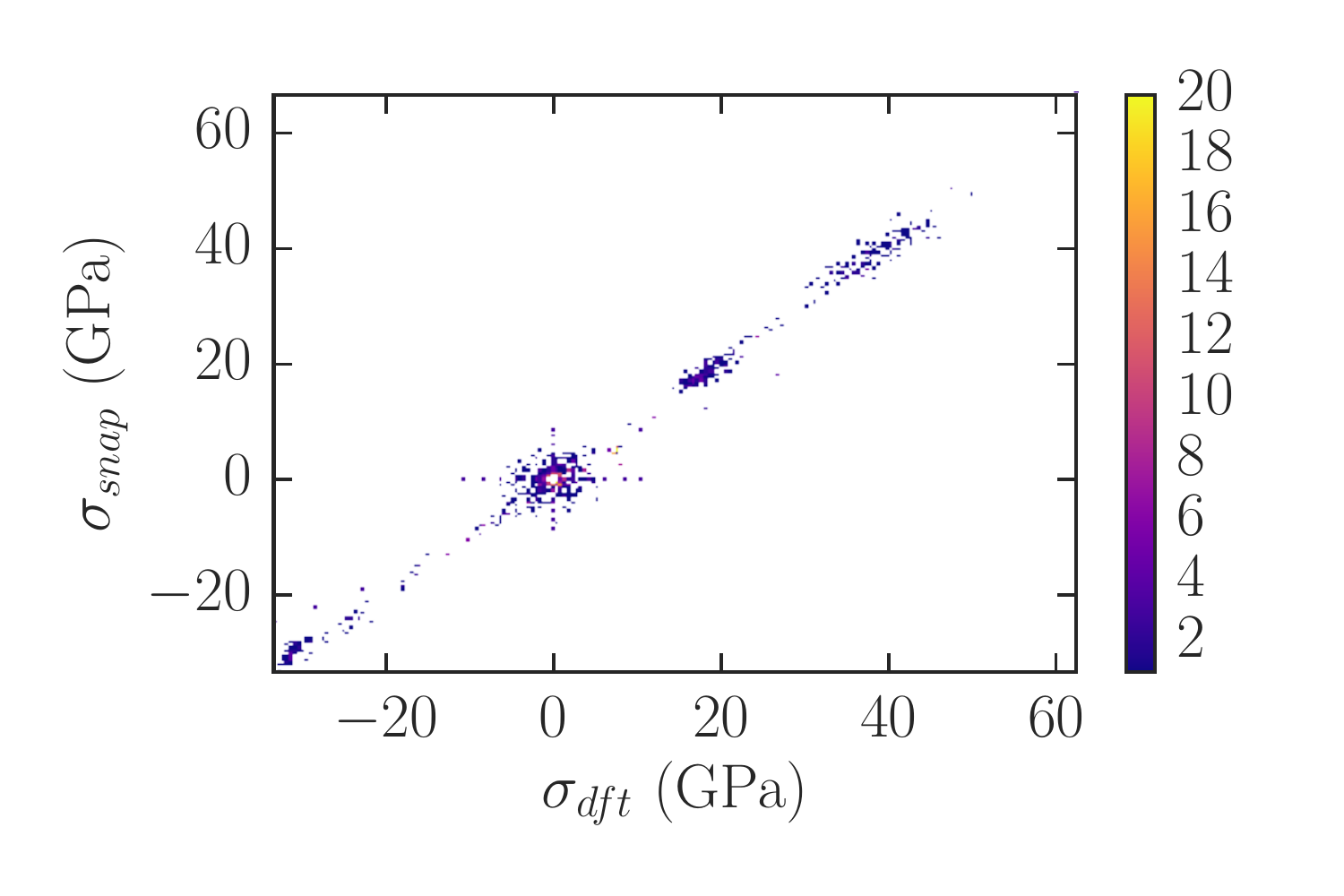}
}
\caption{\label{fig:trainingerror} Histogram of SNAP predictions compared with DFT for (a) energy (b) force and (c) virial stress component.}
\end{center}
\end{figure}

Figure \ref{fig:trainingerror} shows 2D histograms of the comparison between the DFT and SNAP predicted energies, forces and stress components using the training dataset. For all three quantities, the SNAP model predictions are in line with the DFT results with a unity slope. The mean absolute error (MAE) between the DFT and SNAP predictions are 8.9 meV/atom, 0.30 eV/\AA, and 1.26 GPa for the energies, forces and stress components, respectively. In comparison, the corresponding MAEs in the energies, forces, and stress components for the Mo EAM potential of \citet{Zhou2004} are 122 meV/atom, 0.41 eV/\AA ~and 3.88 GPa, respectively, while those for the Mo MEAM potential of \citet{Park2012} are 70 meV/atom, 0.22 eV/\AA, and 1.65 GPa. The Mo SNAP model in this work provides almost an order of magnitude better accuracy in the energy predictions, and good improvement in the accuracy of the stress components. For the prediction of forces, the Mo SNAP model clearly outperforms the EAM potential, but performs slightly worse than the MEAM potential. Nevertheless, it should be noted that the range of the DFT forces spans from around -23 eV/\AA ~to 25 eV/\AA~, and the error of SNAP and MEAM compared to this range is  small. As shown in Figure S1b, the force predictions of MEAM start to deviate substantially when the DFT forces are larger than 10 eV/\AA, while both the SNAP and EAM models maintain a reasonably linear correlation with the DFT calculated forces.

To further validate our model, we performed DFT calculations on the previously excluded vacancy structures and used this dataset of 120 structures as a test case. The predicted MAEs for the energies, forces and stress components are 6.2 meV/atom, 0.27 eV/\AA~and 1.73 GPa respectively, comparable to the model performance on the training datasets. The results confirm that the bispectrum coefficient features are remarkably effective at distinguishing variations in the local environment, and our feature distribution PCA data selection strategy is valid.

\subsection{Lattice constant, elastic constants and equation of state}

Table \ref{dftcomparison} provides a comparison of the Mo SNAP model predictions of the cubic lattice constant and elastic properties of bcc Mo with other force fields and experiments\cite{Simmons1971,Pearson2013,Askill1963}. We find that the calculated cubic lattice constant and elastic properties of the SNAP model are in extremely good agreement with the DFT and experimental values. For example, the SNAP predictions of $c_{11}$, $c_{12}$ and $c_{44}$ are 473 GPa, 158 GPa and 106 GPa respectively, with errors of 0.2\%, 3.8\% and 0\%  compared to DFT, while the errors of the EAM (3.4-8.5\%) and MEAM ($\sim$ 10\%) potentials are significantly higher. The bulk modulus estimated using the Voigt-Reuss-Hill approximation\cite{Hill1952} ($B_{VRH}$) with DFT, SNAP and EAM are in good agreement and slightly lower than the experimental value, but that from the MEAM potential is significantly underestimated.

\begin{table}[htp]
\begin{center}
\caption{Calculated cubic lattice parameter $a$, elastic constants ($c_{ij}$), Voigt-Reuss-Hill bulk modulus ($B_{VRH}$)\cite{Hill1952}., vacancy formation energy ($E_v$) and migration energy ($E_m$) with the DFT, SNAP, EAM and MEAM. Error percentages with respect to DFT values are shown in parentheses.}
\footnotesize
\begin{tabular}{cccccc}
\hline
\hline
& \ \ DFT \ \ & \ \  SNAP \ \ & \ \  EAM \cite{Zhou2004}\ \ & \ \  MEAM\cite{Park2012}  \ \ & Exp. \\
 \hline
 $a$ (\AA)  & 3.168 & 3.160 (-0.3\%) & 3.150 (-0.6\%) & 3.167 (0\%) & 3.147\cite{Pearson2013}\\
 $c_{11}$ (GPa)  & 472 & 473 (0.2\%)& 456 (-3.4\%)& 423 (-10\%) & 479\cite{Simmons1971} \\
 $c_{12} $(GPa)   & 158 & 152 (-3.8\%) & 167 (5.7\%) & 143 (-9.5\%) & 165\cite{Simmons1971} \\
 $c_{44} $(GPa)   & 106 & 106 (0\%) & 115 (8.5\%) & 95 (-10.4\%)& 108\cite{Simmons1971} \\
  $B_{VRH}$ (GPa)   & 263 & 258 (-1.9\%) & 264 (0.4\%) & 236 (-10.3\%) & 270\cite{Simmons1971} \\
$E_v$ (eV) & 2.87 & 2.56 (-10.8\%) & 3.02 (5.2\%) & 2.99 (4.2\%) & - \\
$E_m$ (eV) & 1.12 & 1.39 (24.1\%) & 1.54 (37.5\%) & 1.64(46.4\%) & - \\
$E_a = E_v + E_m$ (eV) & 3.99 & 3.95 (-0.1\%)& 4.56 (14.3\%) & 4.63 (16.0\%)& 4.00 (1850-2350$^\circ$C)\cite{Askill1963} \\
\hline
\hline
\end{tabular}
\label{dftcomparison}
\end{center}
\end{table}

\begin{figure}[!htb]
\includegraphics[width=0.8 \textwidth]{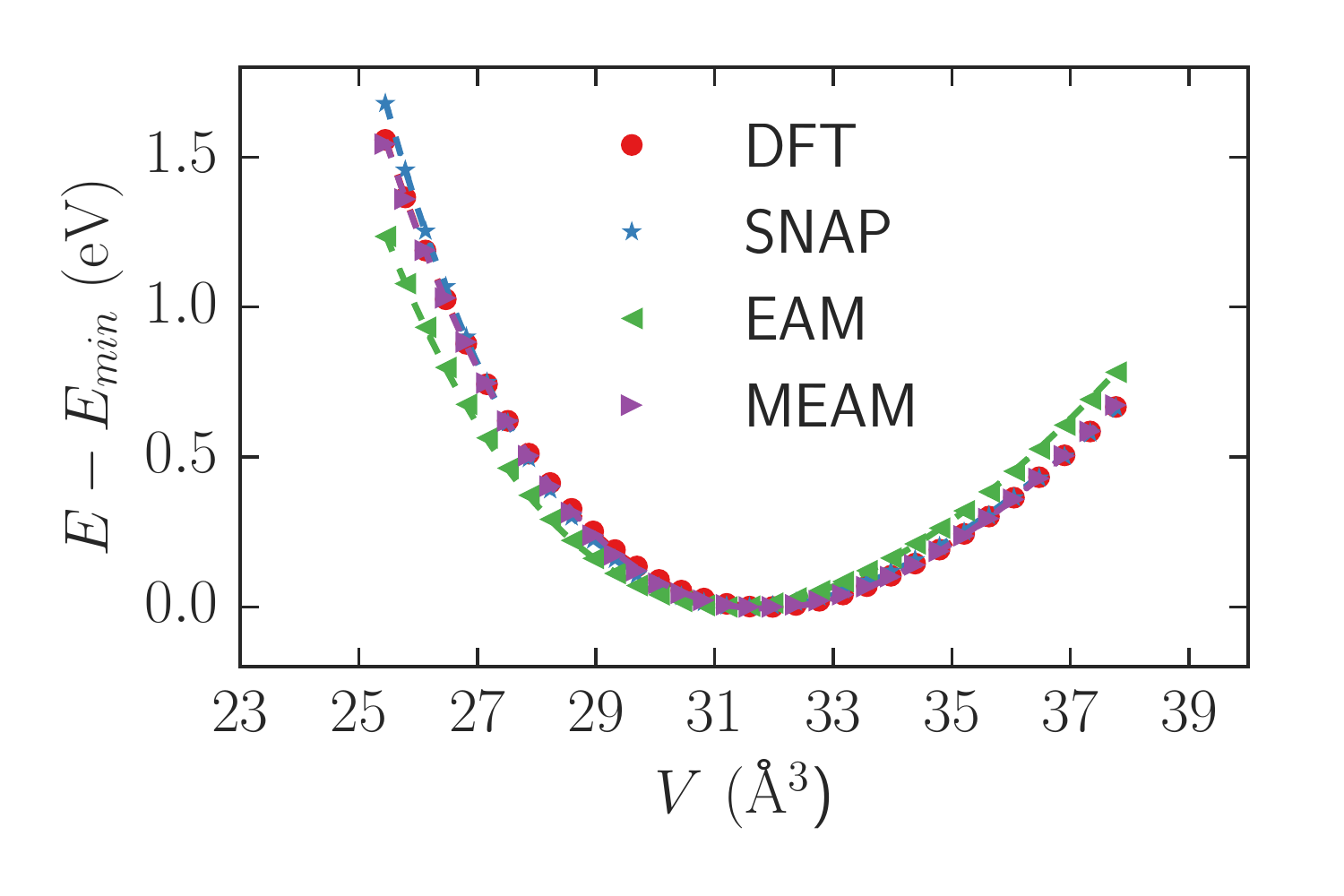}
\caption{\label{fig:BM_equation} Energy versus volume curves of a conventional bcc Mo cell for the DFT, SNAP, EAM and MEAM models. The dash lines show the fitted Murnaghan equation of state. The energy at the equilibrium volume has been set as the zero reference.}
\end{figure}

We have also constructed the energy-versus-volume equation of state curves using DFT, SNAP, EAM and MEAM potentials in Figure \ref{fig:BM_equation}. It should be noted that this set of data was not included in the training data and works as test data for model evaluation. We observe that the SNAP curve overlaps with DFT for volume changes in the range of -15\% to 19\% from the equilibrium volume, but begins to deviate slightly when a larger volume compression of magnitude $>$ 15\% is applied. The EAM potential deviates significantly from the DFT curve at both tensile and compressive strains, while for the MEAM potential, the agreement with DFT is slightly better than the SNAP model. By fitting the Murnaghan equation of state, the estimated bulk moduli from Figure \ref{fig:BM_equation} are 259, 261, 254, and 261 GPa for DFT, SNAP, EAM and MEAM respectively. We note that this estimate of the bulk modulus for the MEAM potential deviates significantly from that estimated using the Voigt-Reuss-Hill approximation (Table \ref{dftcomparison}), and is in much better agreement with the experimental value.

\subsection{Lattice dynamics}

\begin{figure}[!htb]
\includegraphics[width=0.8 \textwidth]{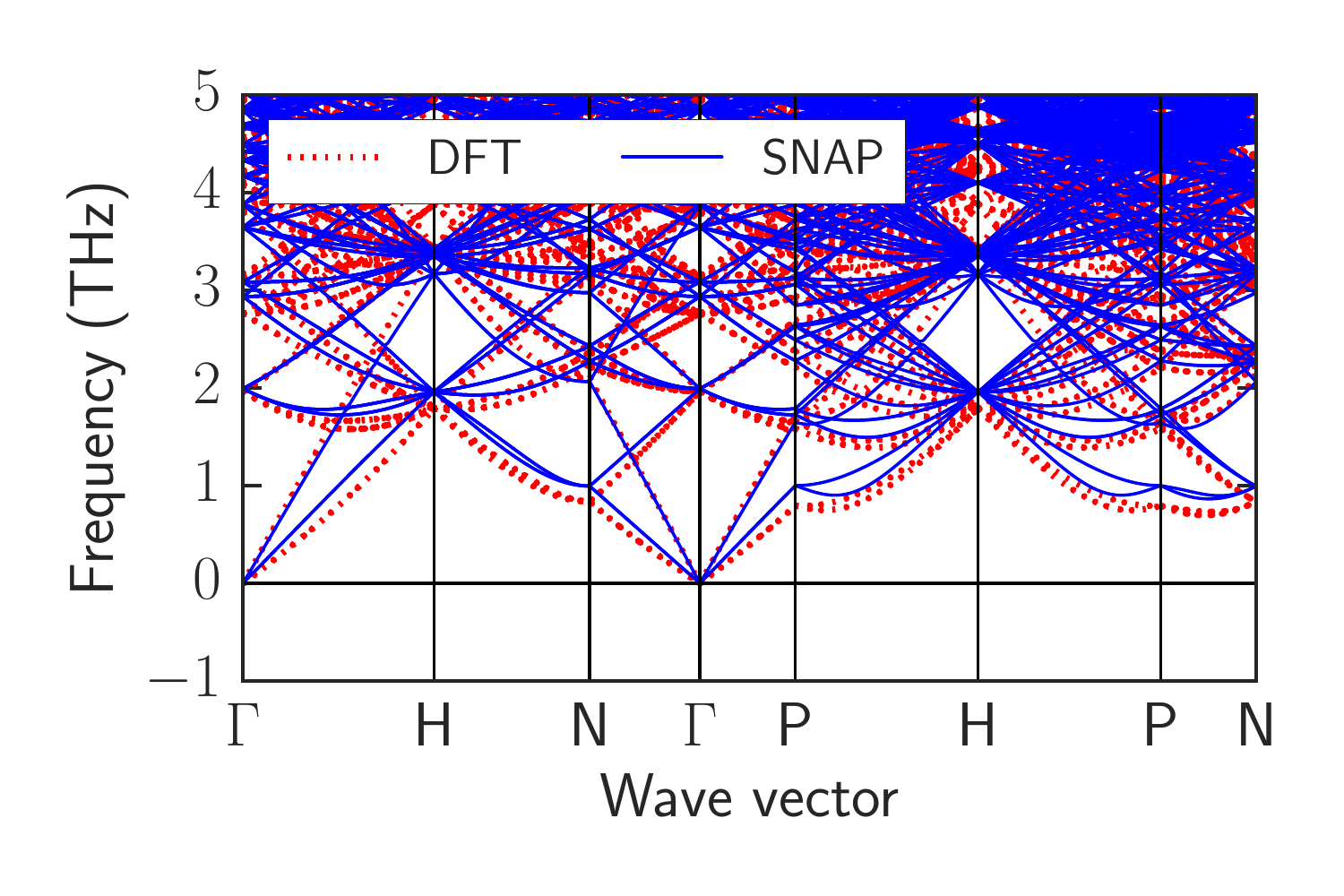}
\caption{\label{fig:phonon} Phonon dispersion curves of Mo SNAP model compared to DFT.}
\end{figure}

\begin{figure}[htp]
\begin{center}
\subfigure[DFT thermal property]{\label{fig:dft_thermal}
\includegraphics[width=.45\textwidth]{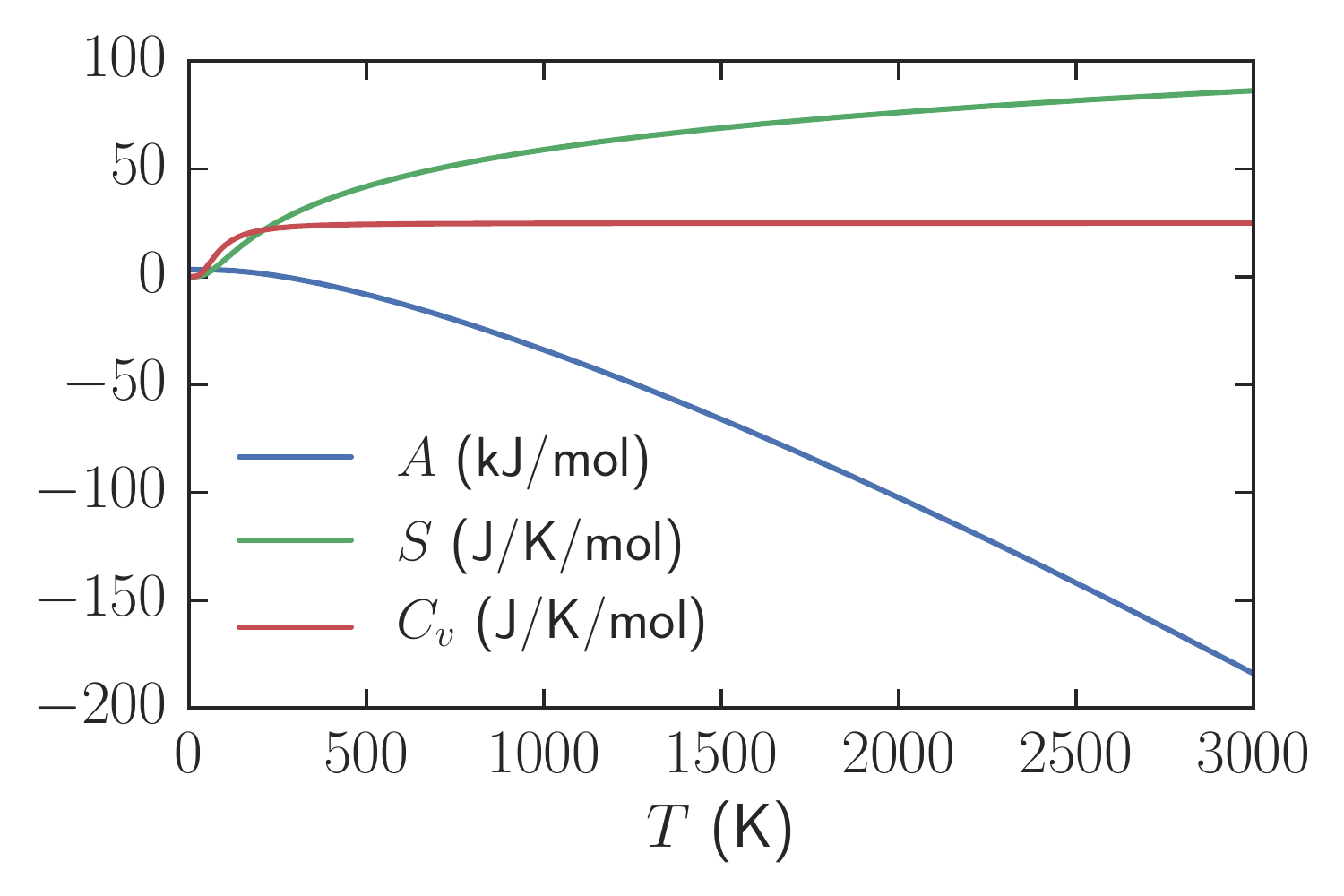}
}
\subfigure[SNAP thermal property]{\label{fig:snap_thermal}
\includegraphics[width=.45\textwidth]{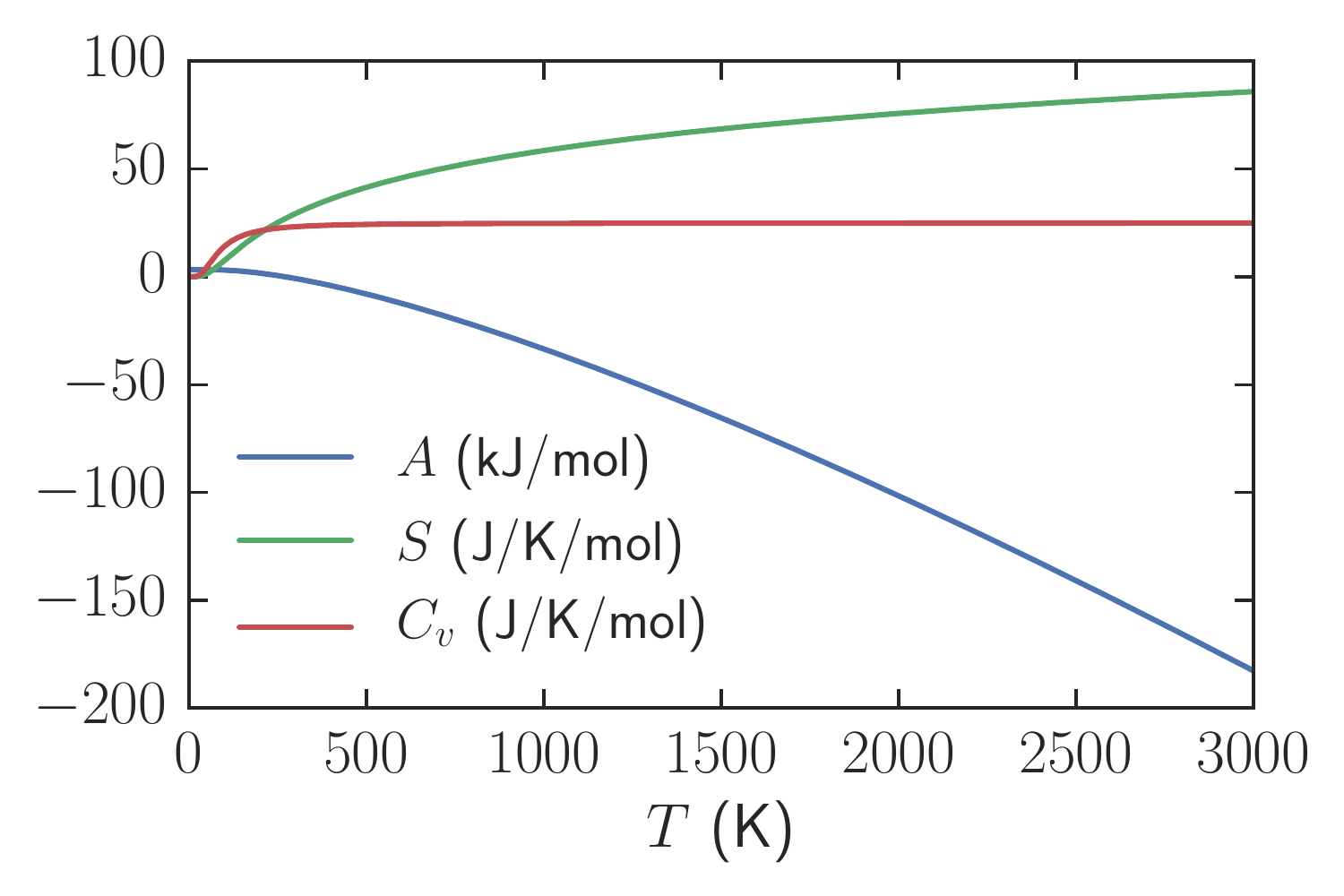}
}

\caption{\label{fig:thermal_property} Helmholtz free energy ($A$), entropy ($S$) and constant volume molar thermal capacity ($C_v$) of Mo calculated with DFT (a) and SNAP (b).}
\end{center}
\end{figure}

To further investigate the prediction of forces and lattice dynamics using the Mo SNAP, the phonon dispersion curve of a Mo $5 \times 5 \times 5$ supercell containing 250 atoms was calculated by feeding the force predictions into the phonopy\cite{Togo2008} package and is shown in Figure \ref{fig:phonon}. The predicted phonon dispersion curves are in good agreement with the DFT calculated phonon dispersion curves, though a systematic slight overestimation (relative to DFT) of the calculated frequencies are seen with the SNAP. No imaginary frequencies are observed, and the lowest frequency lies on the $\Gamma$ point. The phonon dispersion curves of both EAM and MEAM are provided in Figure S2. Both EAM and MEAM show the same overestimation. We further calculated the thermal properties of Mo by employing different models. Figure \ref{fig:thermal_property} shows the Helmholtz free energy ($A$), entropy ($S$) and constant volume molar thermal capacity ($C_v$) of Mo calculated by DFT and SNAP. Both approaches show almost identical curves for all the three quantities. The estimated value of the heat capacity $C_v$ at 300 K is 23.307 J/K/mol for DFT and 23.244 J/K/mol for the SNAP model, an discrepancy of only 0.27\%. The EAM and MEAM potentials also similarly give relatively good predictions of the thermal properties, as shown in Figure S3.

\subsection{Vacancy formation and migration energies}

The formation and migration of defects such as vacancies are of immense interest in practical applications of Mo, especially at high temperatures. Here, we estimate the vacancy formation and migration energies with the climbing-image nudged elastic band (CI-NEB) method\cite{Henkelman2000} as well as molecular dynamics (MD) using the Mo SNAP model.

The DFT, SNAP, EAM and MEAM calculated Mo vacancy formation energy $E_v$ and CI-NEB migration energy $E_m$ are given in Table \ref{dftcomparison}. The SNAP model underestimates $E_v$ by $\sim$ 11\%, while EAM and MEAM overestimate it by 4-5\%. For all force fields, $E_m$ is predicted to be higher than the DFT value, though that predicted by the SNAP model is the closest. The overall predicted activation barrier for vacancy diffusion ($E_a = E_v + E_m$) are 3.99, 3.95, 4.56 and 4.37 eV for the DFT, SNAP, EAM and MEAM models, respectively. Both the DFT and SNAP predicted $E_a$ are close to the experimental activation energy of 4.00 eV (measured on single crystal Mo at the temperature range from 1850-2350 $^\circ$C\cite{Askill1963}).

\begin{figure}[htp]
\begin{center}
\subfigure[MSD]{\label{fig:msd}
\includegraphics[width=.45\textwidth]{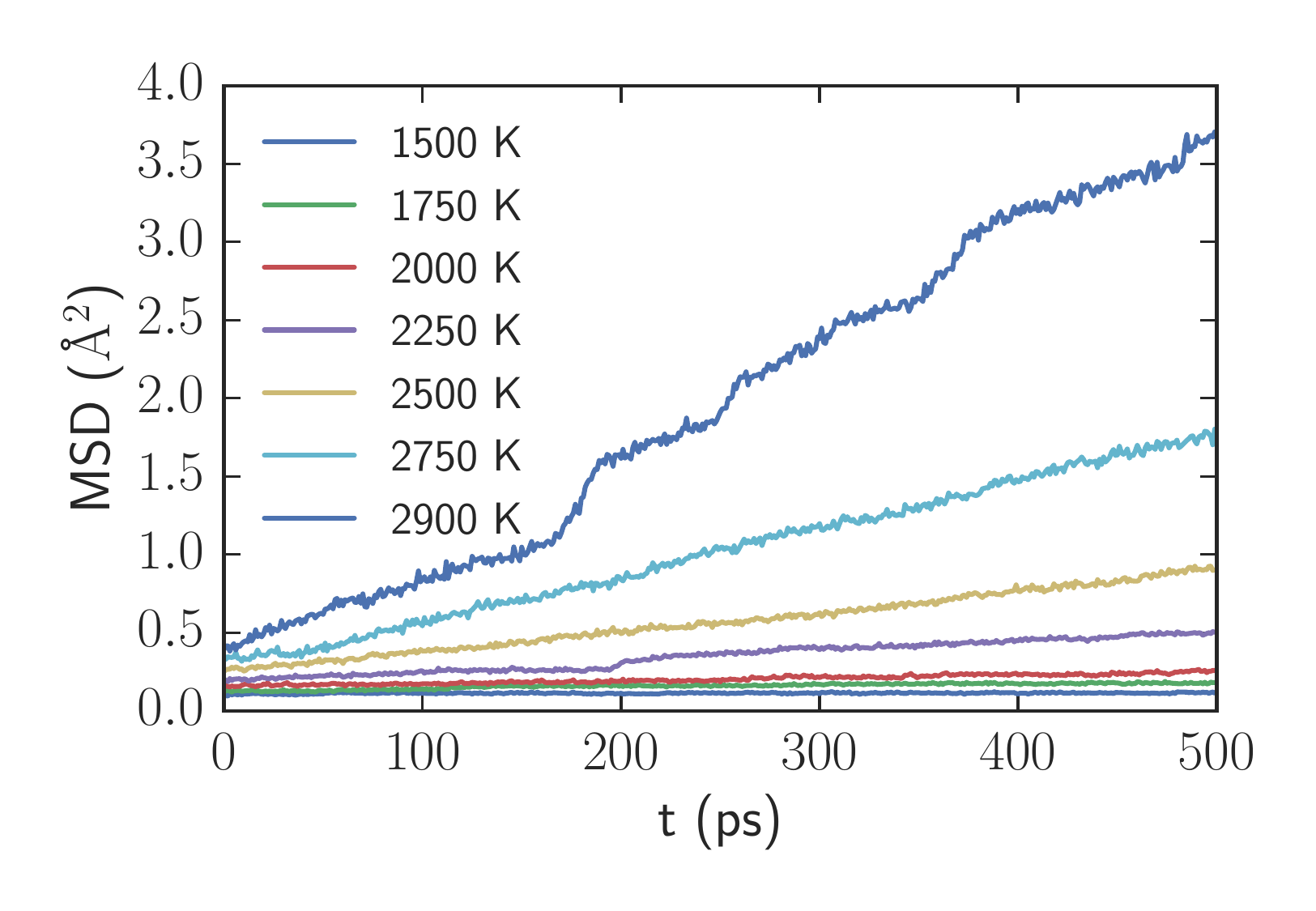}
}
\subfigure[Arrhenius plot]{\label{fig:arrhenius}
\includegraphics[width=.45\textwidth]{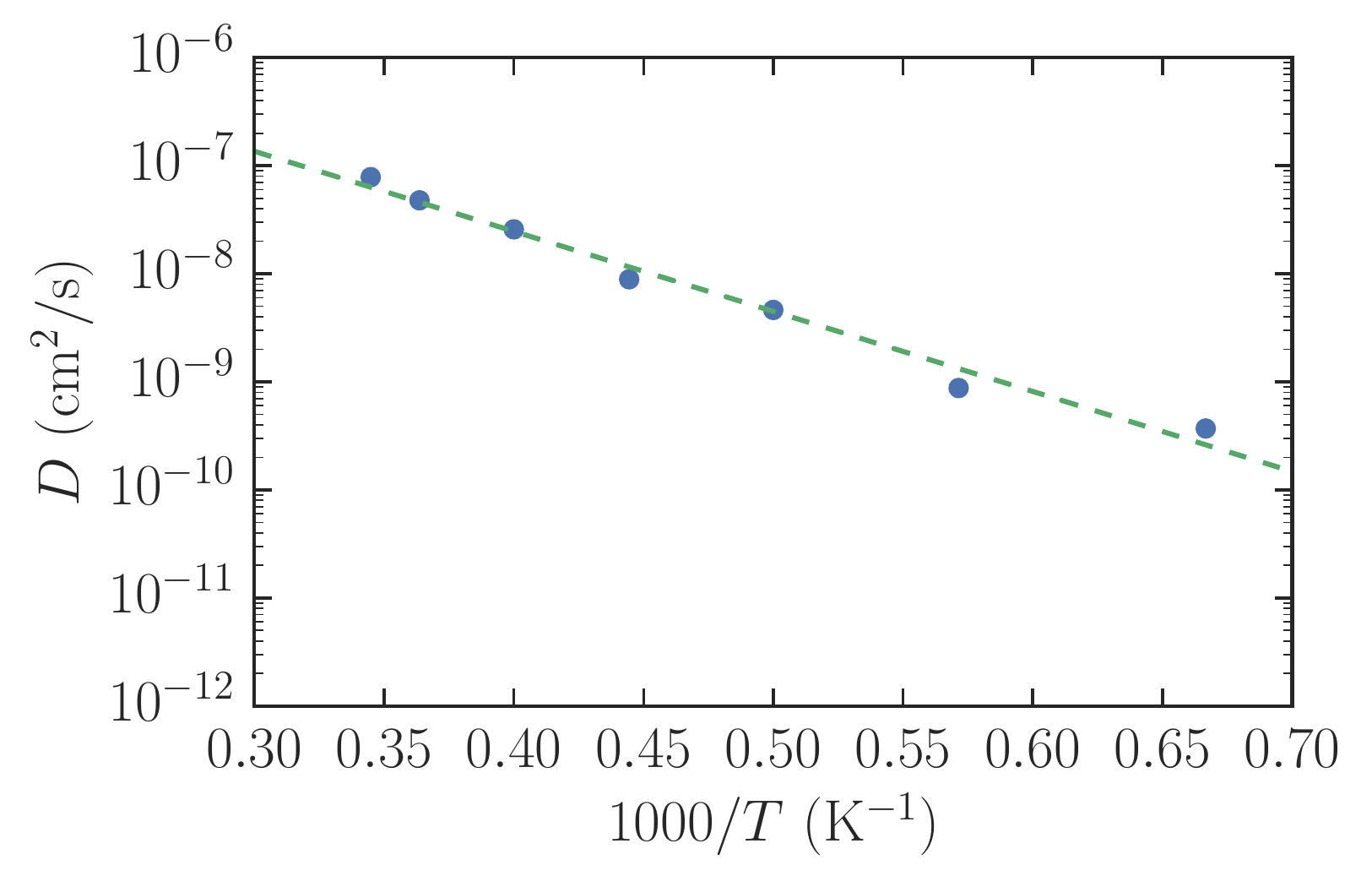}
}

\caption{\label{fig:diffusion} (a) MSDs of a $10\times10\times10$ Mo bcc cell with 1 vacancy (0.05\% concentration) at various temperatures and (b) the corresponding Arrhenius plot.}
\end{center}
\end{figure}

We have also performed MD simulations of a $10\times10\times10$ Mo bcc cell containing one vacancy (1999 atoms with the vacancy concentration of 0.05\%) over 500 ps at seven temperatures (1500-2900 K) using the Mo SNAP.  $NpT$ simulations were carried out using LAMMPS with a time step 1 fs. For each simulation, an equilibration run was carried out over 10 ps, and data was collected during the production run of 500 ps. The mean-squared displacements (MSDs) with respect to simulation time are shown in Figure \ref{fig:msd}. The diffusion of Mo is extremely slow below the melting point, and the calculated self-diffusivity is $\rm 2.57\times10^{-8} cm^2/s$ at 2500 K. This is two orders of magnitude higher than the measured experimental diffusivity of about $\rm 8.22\times10^{-10} cm^2/s$ at 2513 K\cite{Askill1963}. We attribute this discrepancy to the fact that the experimental vacancy concentration may be much lower given the high vacancy formation energy, and is temperature dependent. In the MD simulation, a vacancy was artificially introduced, and the vacancy concentration was fixed at all temperatures. From the Arrhenius plot in Figure \ref{fig:arrhenius}, we estimate the vacancy migration barrier for Mo to be 1.46 eV, which is consistent with the CI-NEB calculated value $E_m$ in Table \ref{dftcomparison}.

\subsection{Melting point}

\begin{figure}[htp]
\includegraphics[width=0.8 \textwidth]{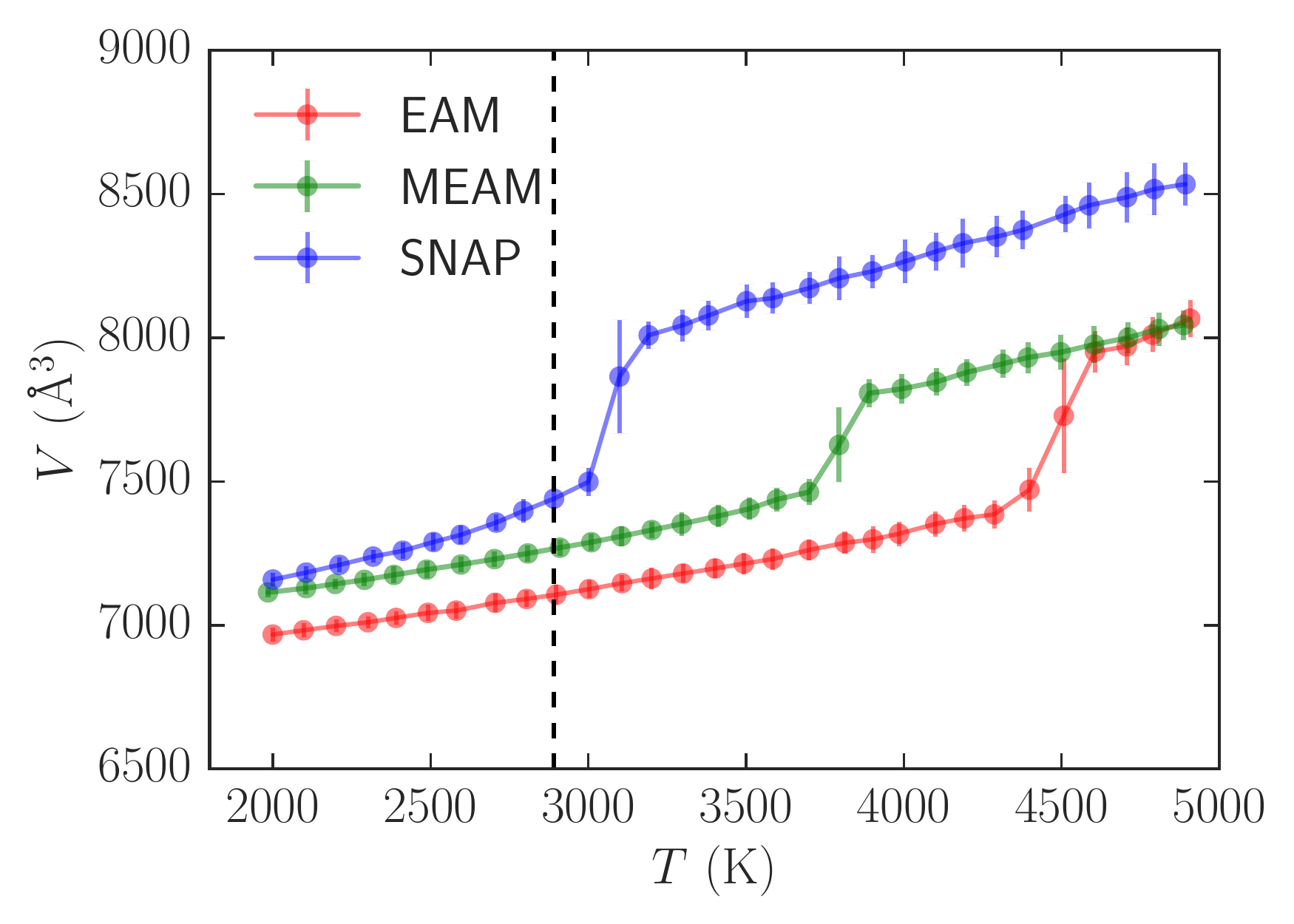}
\caption{\label{fig:melting} Heating simulations of a $6\times6\times6$ Mo cell. The vertical dash line indicates the experimental melting point at 2890.15 K. Error bars of the volume at each temperature are shown.}
\end{figure}

We investigated the ability of the Mo SNAP to reproduce the melting point, which is one of the most challenging properties for conventional force fields to predict. The challenge arises because the phase transition results in a sharp change in the interatomic forces, which are difficult to describe in simple interaction terms. In this work, we performed an $NpT$ MD simulation of a $6\times6\times6$ conventional Mo bcc cell using the SNAP, EAM and MEAM models. The simulation time step was set to 1 fs. The simulations were started at 300 K, and the temperature was ramped up to the desired temperature over 1 ps, followed by equilibration over 10 ps before data collection. Figure \ref{fig:melting} plots the cell volume against the temperature. We find that the SNAP model predicts a melting point of 3000 K, which is in good agreement with the experimental melting point of 2890.15 K. In contrast, both the EAM and MEAM potentials significantly overestimate the melting point of Mo by more than 700 K. We attribute the much better prediction of the SNAP model to the effectiveness of the bispectrum coefficients as a local environment descriptor (whether in the solid or liquid state).

\subsection{Surfaces and grain boundaries}

\begin{figure}[htp]
\includegraphics[width=0.8 \textwidth]{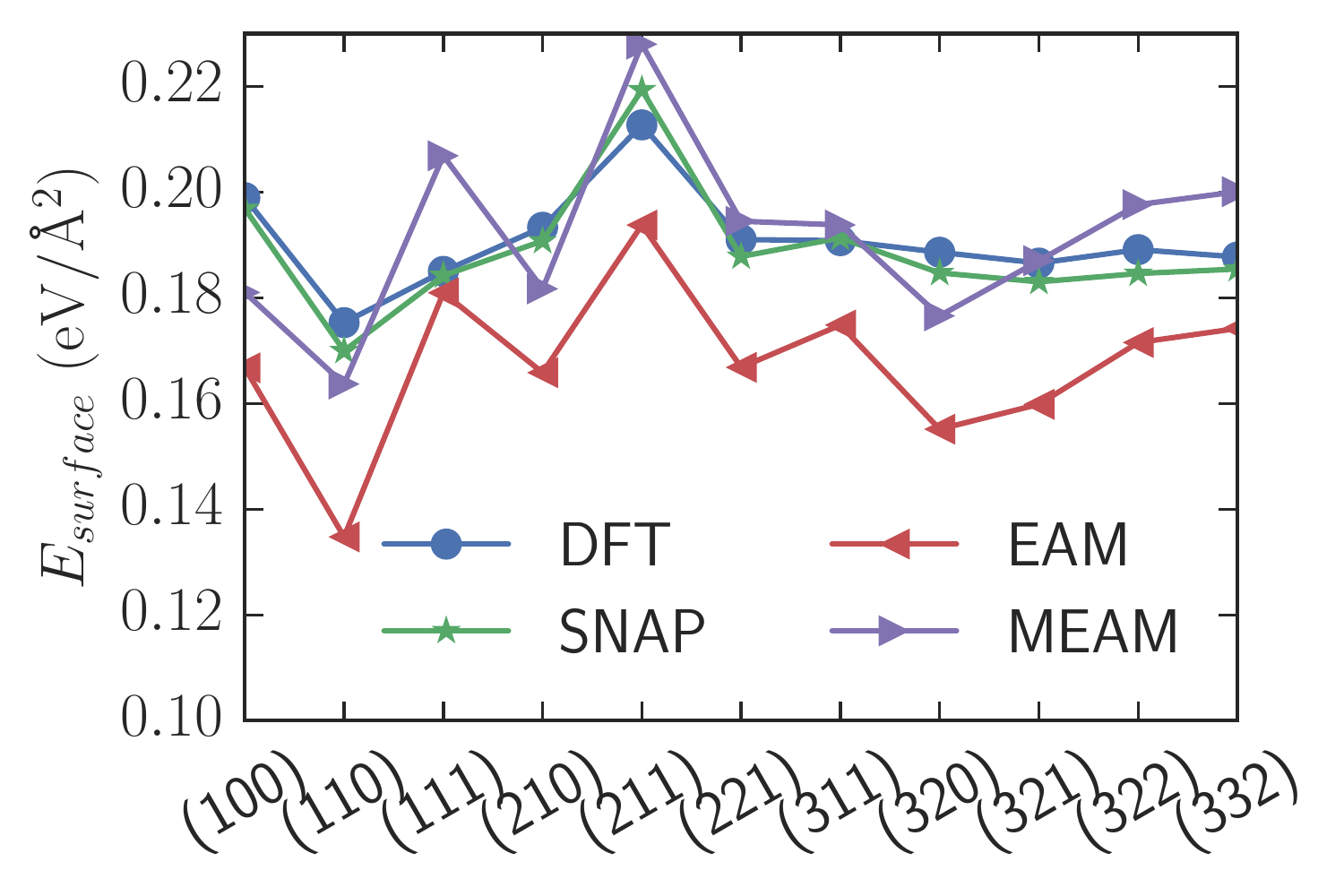}
\caption{\label{fig:surface} Comparison of calculated surface energies for surfaces with Miller indices up to a maximum of 3 using DFT, SNAP, EAM and MEAM.}
\end{figure}

Finally, the performance of the Mo SNAP was assessed with regards to its ability to predict surface and GB energies. As can be seen from Figure \ref{fig:surface}, the surfaces energies computed by the Mo SNAP using the DFT-relaxed slab structures are in excellent agreement with the DFT calculations for both low and high Miller index surfaces. If the slab structures are first relaxed with the SNAP model, the predicted surface energies are somewhat lower than those from DFT calculations. Regardless of whether the slabs are relaxed with the SNAP model, the qualitative trends are reproduced well, as shown in Figure S4. For example, the lowest energy surface is predicted to be (110), with the (111) surface only slightly higher in energy. The errors of EAM and MEAM potentials in predicting the surface energies are much higher. More importantly, the predicted surface energies by the EAM and MEAM potentials are qualitatively different from those of DFT. For example, the predicted surface energies of (111), (322) and (332) are higher than that of (100), which is the opposite from that of DFT calculations.

From the previous work by the authors\cite{Tran2016a}, the calculated DFT energies of the (100) $\Sigma 5$ twist and (310) $\Sigma 5$ tilt GBs are 2.46 J/m$^2$ and 1.81 J/m$^2$, respectively. The optimized SNAP model predicts GB energies of 2.52  J/m$^2$ and 1.94 J/m$^2$ for the (100) $\Sigma 5$ twist and (310) $\Sigma 5$ tilt GBs, which are in good agreement with the DFT values.

\section{Discussion}

\subsection{Limitations of SNAP model}

In this work, we have developed a SNAP model for Mo that significantly outperforms existing EAM and MEAM potentials across a broad range of properties, including energies, structural stresses, elastic constants, thermochemical properties, melting point, surface and grain boundary energies. We attribute this overall better performance to the effectiveness of the bispectrum coefficients as a unique descriptor for the local environment that is invariant to transformations that preserve material properties.

The most notable ``failure'' of the optimized SNAP model is in the prediction of the vacancy formation energy, where the EAM and MEAM potentials perform significantly better (see Table \ref{dftcomparison}). We speculate that this could be due to the fact that the SNAP models the energy as a simple linear relation to the bispectrum descriptors, which undergo a large, discontinuous change when a vacancy is introduced. This is a deficiency that can potentially be addressed by relaxing the linear constraint. On the other hand, the SNAP model provides significantly more accurate vacancy migration energies compared to the EAM and MEAM models. We believe this is because the pairwise interactions in EAM and MEAM models are not able to capture the transition state with sufficient accuracy. The overall errors in the vacancy formation and migration energies cancel out, resulting in overall activation energies that are remarkably close to experimental values.

Another disadvantage of the SNAP model is its higher ($\sim 2-3$ orders of magnitude) computational cost relative to MEAM. Nevertheless, the SNAP model is still orders of magnitude cheaper than DFT calculations and have the advantage of approximately linear scaling with respect to number of atoms\cite{Thompson2015}. Using a single computing node of 24 cores, we were able to carry out a MD simulation of $\sim 2000$ atoms for hundreds of ps ($> 100,000$ timesteps of 1 fs) over 24 hours. Simulations of tens of thousands of atoms should be well within the capabilities of modern supercomputing clusters. Access to such length and time scales with close to DFT chemical accuracy would significantly enhance our ability to probe interesting science in lower symmetry systems, for example, crack propagation and effect of dopants in high $\Sigma$ or general grain boundaries.

\subsection{Model development}

We have also demonstrated several enhancements that we believe would be of relevance to future efforts in the development of machine-learned potentials.

First, we have leveraged on a combination of existing data from previous works\cite{Jain2013, Tran2016, crystalium2016, Tran2016a}, as well as newly generated structures from high-throughput DFT calculations for model training and testing. The use of well-validated, pre-existing data led to a significantly more streamlined and efficient model development process. The pre-existing data in this work are mostly from previous works by the current authors, some of which (e.g., surface energies) are in large, open access online databases.\cite{Tran2016, crystalium2016, materialsproject} We foresee that the continued proliferation of such central materials data repositories would provide future potential developers with a wealth of data for model training and testing. We note that a key gap is that existing open databases such as the Materials Project\cite{Jain2013} are focused mainly on ground state structures, energies and properties (e.g., elastic constants).  To our knowledge, there are currently no open databases for trajectory data from AIMD simulations, which has been shown to be an extremely rich data set for model training. Together with this work, we have published the trajectory data for the Mo AIMD simulations (as well as all other data used in model development and the final data weights and optimized SNAP model parameters) in an open repository hosted on Github (https://github.com/materialsvirtuallab/snap). We hope to address this gap through a more systematic large-scale effort encompassing more diverse chemistries in future.

Second, we have shown that an exploratory data analysis performed prior to expensive DFT calculations can result in better quality training data with lower computational effort. In this work, we utilized a PCA approach to identify the datasets that provide distinct local environment information. This data selection process avoids biasing the training with duplicate data that share common features and also improves the accuracy of the eventual model. For instance, we have previously excluded the vacancy data group based on its overlap with the AIMD groups.

Finally, similar to previous works, we find that the hyperparameters, i.e., the cut-off radius $R_c$ for the bispectrum calculations and weights for different data group, can influence the model performance and have to be optimized together with the model training. The choice of the cut-off radius $R_c$ has a substantial effect on model accuracy (e.g., in energy and force predictions), while the choice of data weights tend to impact derived material properties. For example, the $c_{12}$ elastic constant can vary from 20 GPa to 400 GPa with changes in data weights. A global optimization approach, such as the differential evolution algorithm used in this work, can improve the likelihood of obtaining good solutions, if the bounds are properly set.

\section{Conclusion}

To conclude, we have developed a SNAP model for Mo by applying a systematic data selection and global optimization approach on a combination of existing as well as newly generated data from DFT calculations. The optimized SNAP model outperforms existing EAM and MEAM potentials for Mo, achieving close to DFT accuracy in the prediction of a spectrum of properties of immense fundamental and technological relevant, including energies, forces, elastic constants, melting point, surface and grain boundary energies. We believe this new chemically-accurate and efficient Mo SNAP potential open ups our ability to probe interesting science with simulations of large-scale models over longer time scales.

\section{acknowledgement}
This  work  was  supported  by  the  Office of Naval Research, under Contract No. N000141612621.  The  authors  also  acknowledge computational resources provided by Triton Shared Computing Cluster (TSCC) at the University of California, San Diego, the National Energy Research Scientific Computing Center (NERSC), and the Extreme Science and Engineering Discovery Environment (XSEDE) supported by National Science Foundation under grant number ACI-1053575.

\clearpage
\bibliography{library}
\clearpage

\end{document}